\documentclass[conference]{IEEEtran}
\usepackage{graphicx}
\usepackage[utf8]{inputenc}
\usepackage{lmodern}
\PassOptionsToPackage{hyphens}{url}
\usepackage{bbold}

\title{A Practical Quantum Instruction Set Architecture}
\author{\IEEEauthorblockN{Robert S.\ Smith, Michael J.\ Curtis, William J.\ Zeng}
\IEEEauthorblockA{Rigetti Computing\\
775 Heinz Ave.\\
Berkeley, California 94710\\
Email: \{\texttt{robert}, \texttt{spike}, \texttt{will}\}\texttt{@rigetti.com}}}
\date{\today}

\usepackage{amsmath,amsfonts,amssymb,amsthm}
\usepackage[square,sort,comma,numbers]{natbib}
\usepackage{graphicx}
\usepackage{braket}
\usepackage{mathtools}
\usepackage{verbatim}
\usepackage{mathrsfs}
\let\labelindent\relax
\usepackage[inline]{enumitem}
\usepackage{tikz}

\newcommand{\defn}[1]{\textbf{#1}}
\newcommand{\Liquid}{$\textrm{LIQ}Ui\vert\rangle$}

\newcommand{\SWAP}{\ensuremath{\mathsf{SWAP}}}
\newcommand{\CPHASE}{\ensuremath{\mathsf{CPHASE}}}
\newcommand{\RX}{\ensuremath{\mathsf{R}_x}}
\newcommand{\RZ}{\ensuremath{\mathsf{R}_z}}
\newcommand{\HADAMARD}{\ensuremath{\mathsf{H}}}
\newcommand{\CNOT}{\ensuremath{\mathsf{CNOT}}}

\newcommand{\Hil}{\ensuremath{\mathscr{H}}}
\newcommand{\BHil}{\ensuremath{\mathscr{B}}}
\newcommand{\Id}{\mathsf{I}}

\newcommand{\diag}{\mathop{\mathrm{diag}}\nolimits}

\newtheorem{example}{Example}

\newcommand\copyrighttext{%
  \footnotesize Copyright \textcopyright\ 2016 Rigetti \& Co.,\ Inc.---v2.0.20170217}
\newcommand\copyrightnotice{%
\begin{tikzpicture}[remember picture,overlay]
\node[anchor=south,yshift=10pt] at (current page.south) {\fbox{\parbox{\dimexpr\textwidth-\fboxsep-\fboxrule\relax}{\copyrighttext}}};
\end{tikzpicture}%
}

\begin{document}
\maketitle

\copyrightnotice

\begin{abstract}
We introduce an abstract machine architecture for classical/quantum computations---including compilation---along with a quantum instruction language called Quil for explicitly writing these computations. With this formalism, we discuss concrete implementations of the machine and non-trivial algorithms targeting them. The introduction of this machine dovetails with ongoing development of quantum computing technology, and makes possible portable descriptions of recent classical/quantum algorithms.
\end{abstract}

\tableofcontents

\section{Introduction}

The underlying hardware for quantum computing has advanced rapidly in recent years. Superconducting chips with 4--9 qubits have been demonstrated with the performance required to run quantum simulation algorithms~\cite{o2015scalable, geller2015universal, barends2015digitized}, quantum machine learning~\cite{riste2015demonstration}, and quantum error correction benchmarks~\cite{chow2015characterizing, kelly2015state, riste2015detecting}. 

Hybrid classical/quantum algorithms---including variational quantum eigensolvers~\cite{peruzzo2014variational, wecker2015progress, mcclean2015theory}, correlated material simulations~\cite{bauer2015hybrid}, and approximate optimization~\cite{farhi2014quantum}---have much promise in reducing the overhead required for valuable applications.  In machine learning and quantum simulation, particularly for catalysts~\cite{reiher2016elucidating} and high-temperature superconductivity~\cite{wecker2015progress}, scalable quantum computers promise performance unrivaled by classical supercomputers.

The promise of hardware and applications must be matched with advances in programming architectures. The demands of practical algorithm design, as well as the shift to hybrid classical/quantum algorithms, necessitate an update to the \emph{quantum Turing machine} model~\cite{Deutsch97}. We need new frameworks for quantum program compilation~\cite{Liquid, Quipper, haner2016software, svore2006layered, javadiabhari2015scaffcc} and emulation~\cite{haner2016high, smelyanskiy2016qhipster}. For more details on prior work and its relationship to the topics introduced here, we refer the reader to Appendix~\ref{sec:prior}.

\begin{figure}[t]
    \centering
    \includegraphics[scale=0.5]{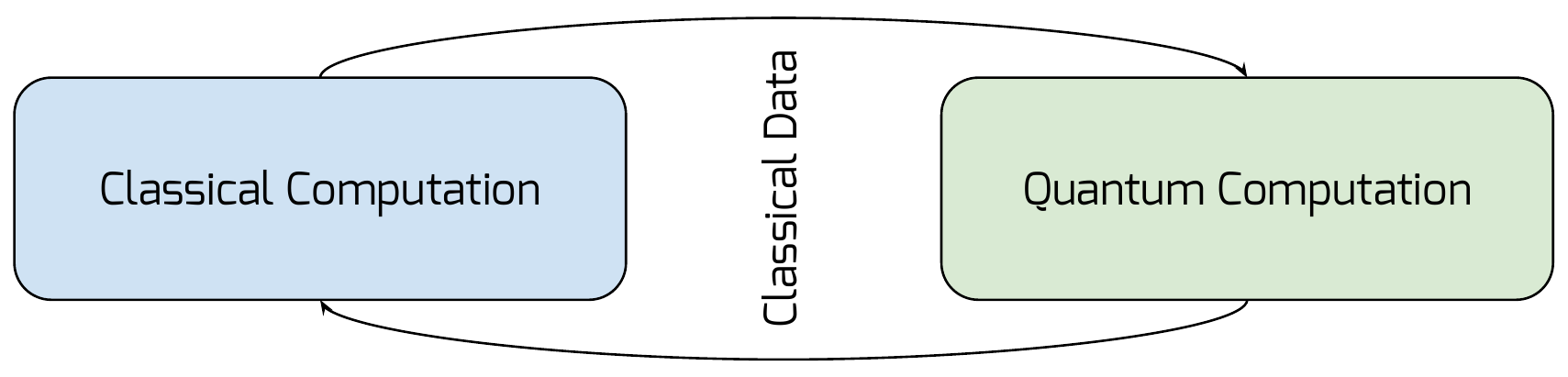}
    \caption{A classical/quantum feedback loop.}
    \label{fig:classical-quantum-feedback}
\end{figure}

These classical/quantum algorithms require a classical computer and quantum computer to communicate and work cooperatively, as in Figure~\ref{fig:classical-quantum-feedback}. Within the presented framework, classical information is fed back via a defined memory model, which can be implemented efficiently in both hardware and software.

In this paper, we describe an abstract machine which serves as a model for hybrid classical/quantum computation. Alongside this, we describe an instruction language for this machine called Quil and its suitability for program analysis and compilation. Together these form a \defn{quantum instruction set architecture} (ISA). Lastly, we give various examples of algorithms in Quil, and discuss an executable implementation.

\section{The Quantum Abstract Machine}\label{sec:qam}
Turing machines provide a vehicle for studying important concepts in computer science such as complexity and computational equivalence. While theoretically important, they do not provide a foundation for the construction of practical computing machines. Instead, specialized abstract machines are designed to accomplish real-world tasks, like arithmetic, efficiently while maintaining Turing completeness. These machines are often specified in the form of an instruction set architecture. Quantum Turing machines lie in the same vein as its classical counterpart, and we follow a similar approach in the creation of a practical quantum analog.

The \defn{Quantum Abstract Machine} (QAM) is an abstract representation of a general-purpose quantum computing device. It includes support for manipulating both classical and quantum state. The QAM executes programs represented in a quantum instruction language called \defn{Quil}, which has well-defined semantics in the context of the QAM.

The state of the QAM is specified by the following elements:
\begin{itemize}
\item A fixed but arbitrary number of qubits $N_q$ indexed from $0$ to $N_q-1$. The $k$\textsuperscript{th} qubit is written $Q_k$. The state of these qubits is written $\ket{\Psi}$ and is initialized to $\ket{00\ldots0}$. The semantics of the qubits are described in Section~\ref{sec:qubitsemantics}.
\item A classical memory $C$ of $N_c$ bits, initialized to zero and indexed from $0$ to $N_c-1$. The $k$\textsuperscript{th} bit is written $C[k]$.
\item A fixed but arbitrary list of static gates $G$, and a fixed but arbitrary list of parametric gates $G'$. These terms are defined in Section~\ref{sec:gates}.
\item A fixed but arbitrary sequence of Quil instructions $P$. These instructions will be described in Section~\ref{sec:quil}.
\item An integer program counter $0 \le \kappa \le \vert P\vert$ indicating position of the next instruction to execute when $\kappa\neq\vert P\vert$ or a halted program when $\kappa=\vert P\vert$.
\end{itemize}
The 6-tuple $(\ket{\Psi}, C, G, G', P, \kappa)$ summarizes the state of the QAM.

The QAM may be implemented either classically or on quantum hardware. A classical implementation is called a \defn{Quantum Virtual Machine} (QVM). We describe one such implementation in Section~\ref{sec:qvm}. An implementation on quantum hardware is called a \defn{Quantum Processing Unit} (QPU).

The semantics of the quantum state and operations on it are described in the language of tensor products of Hilbert spaces and linear maps between them. The following subsections give these semantics in meticulous detail. Readers with intuition about these topics are encouraged to skip to Section~\ref{sec:quil} for a description of Quil.

\subsection{Qubit Semantics}\label{sec:qubitsemantics}
A finite-dimensional Hilbert space over the complex numbers $\mathbb{C}$ is denoted by \Hil. The state space of a qubit is a two-dimensional Hilbert space over $\mathbb{C}$ and is denoted by \BHil. Each of these Hilbert spaces is equipped with a chosen orthonormal basis and indexing map on that basis. An \defn{indexing map} is a bijective function that maps elements of a finite set $\Sigma$ to the set of non-negative integers below $\vert\Sigma\vert$, denoted $[\vert\Sigma\vert]$. For a Hilbert space spanned by $\{\ket{u}, \ket{v}\}$ with an indexing map defined by $\ket{u}\mapsto 0$ and $\ket{v}\mapsto 1$, we write $\ket{0}\coloneqq\ket{u}$ and $\ket{1}\coloneqq\ket{v}$.

In the context of the QAM, each qubit $Q_k$ in isolation has a state space $\BHil_k$ spanned by an orthonormal basis $\{\ket{0}_k,\ket{1}_k\}$ called the \defn{computational basis}. Since qubits can entangle, the state space of the system of all qubits is not a direct product of each constituent space, but rather a rightward tensor product
\begin{equation}\label{eq:tensoredup}
\ket{\Psi} \in \Hil \coloneqq \bigotimes_{k=0}^{N_q-1} \BHil_{N_q - k - 1}.
\end{equation}
The meaning of the tensor product is as follows. Let $\ket{p}_i$ be the $p$\textsuperscript{th} basis vector of $\Hil_i$ according to its indexing map. The tensor product of two Hilbert spaces is then
\begin{equation}\label{eq:hilbertproduct}
\Hil_i \otimes \Hil_j \coloneqq \Big\{\sum_{\substack{p\in[\dim \Hil_i]\\q\in[\dim \Hil_j]}}\!\!\!\!\!\!\mathsf{C}_{p,q}\underbrace{\ket{p}_i\otimes\ket{q}_j}_{\mathclap{\text{basis element}}} \mathbin{:}\, \mathsf{C} \in \mathbb{C}^{\dim \Hil_i \times \dim \Hil_j}\Big\}.
\end{equation}
The resulting basis elements are ordered by way of the \defn{lexicographic indexing map}
\begin{equation}
\ket{p}_i\otimes\ket{q}_j \mapsto q + p\dim \Hil_j.
\end{equation}
Having the basis elements of the Hilbert space $\Hil_i$ ``dominate'' the indexing map is a convention due to the standard definition of the Kronecker product, in which for matrices $A$ and $B$, $A\otimes B$ is a block matrix $(A\otimes B)_{i,j} = A_{i,j} B$. This convention, while standard, somewhat muddles the semantics below with busy-looking variable indexes which count down, not up.

A basis element of $\Hil$ \[\ket{b_{N_q-1}}_{N_q-1}\otimes\cdots\otimes\ket{b_1}_1\otimes\ket{b_0}_0\] can be written shorthand in \defn{bit string notation} \[\ket{b_{N_q-1}\ldots b_1 b_0}.\] This has the particularly useful property that the bit string corresponds to the binary representation of the index of that basis element. For clarity, however, we will not use this notation elsewhere in this paper.

\begin{example}
A two-qubit system in the Hilbert space $\BHil_2\otimes \BHil_1$ has the lexicographic indexing map defined by
\begin{align*}
\ket{0}_2\otimes\ket{0}_1 &\mapsto 0, &
\ket{0}_2\otimes\ket{1}_1 &\mapsto 1,\\
\ket{1}_2\otimes\ket{0}_1 &\mapsto 2, &
\ket{1}_2\otimes\ket{1}_1 &\mapsto 3.
\end{align*}
The standard Bell state in this system is represented by the element in the tensor space with the matrix
\begin{equation*}
\mathsf{C} = \frac{1}{\sqrt{2}}
\begin{pmatrix}
1 & 0 \\ 
0 & 1
\end{pmatrix}.
\end{equation*}
\end{example}

Eqs.~\eqref{eq:tensoredup} and \eqref{eq:hilbertproduct} imply that $\dim \Hil = 2^{N_q}$, and as such, $\ket{\Psi}$ can be represented as a complex vector of that length.

\subsection{Quantum Gate Semantics}\label{sec:gatesemantics}
The quantum state of the system evolves by applying a sequence of operators called \defn{quantum gates}. Most generally, these are complex unitary matrices of size $2^{N_q} \times 2^{N_q}$, written succinctly as $\mathrm{U}(2^{N_q})$. However, quantum gates are typically abbreviated as one- or two-qubit unitary matrices, and go through a process of ``tensoring up'' before application to the quantum state. 

\begin{example}
The \defn{Hadamard gate} on $Q_k$ is defined as
\begin{equation*}
\mathsf{H} \coloneqq \frac{1}{\sqrt{2}}
\begin{pmatrix}
1 & 1\\
1 & -1
\end{pmatrix} : \BHil_k\to\BHil_k.
\end{equation*}
This is unitary because $\mathsf{H}\mathsf{H}^{\dagger} = \Id_k$ is the identity map on $\BHil_k$.

The \defn{controlled-$\mathsf{X}$} or \defn{controlled-not gate} with \emph{control qubit} $Q_j$ and \emph{target qubit} $Q_k$ is defined as
\begin{equation*}
\CNOT \coloneqq
\begin{pmatrix}
1 & 0 & 0 & 0\\
0 & 1 & 0 & 0\\
0 & 0 & 0 & 1\\
0 & 0 & 1 & 0
\end{pmatrix} : \BHil_j\otimes\BHil_k\to\BHil_j\otimes\BHil_k.
\end{equation*}
This gate is common for constructing \emph{entanglement} in a quantum system.
\end{example}

It turns out that many different sets of these one- and two-qubit gates are sufficient for universal quantum computation, i.e., a discrete set of gates can be used to approximate any unitary matrix to arbitrary accuracy~\cite{barenco1995elementary, nielsen2010quantum}. In fact almost any two-qubit quantum gate can be shown to be universal~\cite{deutsch1995universality}. Delving into different universal gate sets is beyond the scope of this work and we refer the reader to~\cite{nielsen2010quantum} as a general reference. 

We now wish to provide a constructive method for interpreting operators on a portion of a Hilbert space as operators on the Hilbert space itself. In the simplest case, we have a one-qubit gate $U$ acting on qubit $Q_k$, which induces an operator $\tilde U$ on $\Hil$ by tensor-multiplying with the identity map a total of $N_q-1$ times:
\begin{equation}\label{eq:tensorone}
\tilde U = \Id_{N_q-1} \otimes \cdots \otimes \underbrace{U}_{\mathclap{\text{$(N_q-k-1)$\textsuperscript{th} position (zero-based)}}} \otimes \cdots \otimes \Id_1 \otimes \Id_0.
\end{equation}
We refer to this process as \defn{lifting} and reserve the tilde over the operator name to indicate such.

\begin{example}
Consider a system of four qubits and consider a Hadamard gate acting on $Q_2$. Lifting this operator according to \eqref{eq:tensorone} gives
\begin{equation*}
\tilde{\mathsf{H}} = \Id_3 \otimes \mathsf{H} \otimes \Id_1 \otimes \Id_0.
\end{equation*}
\end{example}

A two-qubit gate acting on adjacent Hilbert spaces is just as simple. An operator $U:\BHil_{k}\otimes \BHil_{k-1}\to \BHil_k\otimes \BHil_{k-1}$ is lifted as
\begin{equation}\label{eq:tensortwo}
\tilde U = \underbrace{\Id_{N_q-1} \otimes \cdots \otimes U \otimes \cdots \otimes \Id_1\otimes \Id_0}_{\text{$N_q-1$ factors, $U$ at position $N_q-k-1$}}.
\end{equation}
However, when the Hilbert spaces are not adjacent, lifting involves a bit more bookkeeping because there is no obvious place to tensor-multiply the identity maps of the Hilbert spaces indexed between $j$ and $k$. We can resolve this by suitably rearranging the space. We need two tools: a method to ``reorganize'' an operator's action on a tensor product of Hilbert spaces and isomorphisms between these Hilbert spaces. The general principle to be employed is to find some operator $\pi : \Hil \to \Hil$ which acts as a permutation operator on $\Hil$, and to compute $\pi^{-1} U' \pi$, where $U'$ is isomorphic to $U$. Simply speaking, $\pi$ is a temporary re-indexing of basis vectors and $U'$ is just a trivial reinterpretation of $U$. 

To reorganize, we use the fact that any permutation can be decomposed into adjacent transpositions. For swapping two qubits, consider the gate
\begin{equation}
\SWAP_{j,k} \coloneqq
\begin{pmatrix}
1 & 0 & 0 & 0\\
0 & 0 & 1 & 0\\
0 & 1 & 0 & 0\\
0 & 0 & 0 & 1
\end{pmatrix}
: \BHil_j\otimes \BHil_k \to \BHil_j\otimes \BHil_k.
\end{equation}
This is a permutation matrix which maps the basis elements according to
\begin{align*}
\ket{0}_j\otimes\ket{1}_k &\mapsto \ket{1}_j\otimes\ket{0}_k\\
\ket{1}_j\otimes\ket{0}_k &\mapsto \ket{0}_j\otimes\ket{1}_k,
\end{align*}
and mapping the others identically. For adjacent transpositions in the full Hilbert space, this can be trivially lifted:
\begin{equation}
\tau_i \coloneqq \text{$\SWAP_{i,i+1}$ lifted by way of \eqref{eq:tensortwo}.}
\end{equation}
We will use a sequence of these operators to arrange for $\BHil_j$ and $\BHil_k$ to be adjacent by moving one of them next to the other. There are two cases we need to be concerned about: $j > k$ and $j < k$.

For the $j > k$ case, we want to map the state of $\BHil_k$ to $\BHil_{j-1}$. This is accomplished with the operator\footnote{This is an example of the effects of following the Kronecker convention. This product is really just ``$\tau_i$ in reverse.''}
\begin{equation*} 
\pi_{j,k} \coloneqq \prod_{i=k}^{j-2}\tau_{j+k-i-2},
\end{equation*}
where the product right-multiplies and is empty when $k\ge j-1$.

For the $j < k$ case, we want to map the state of $\BHil_{j}$ to $\BHil_{k-1}$, and then swap $\BHil_{k-1}$ with $\BHil_k$. We can do this with the $\pi$ operator succinctly:
\begin{equation*} 
\pi_{j,k}' \coloneqq \tau_{k-1}\pi_{k,j}.
\end{equation*}
Note the order of $j$ and $k$ in the $\pi$ factor.

Lastly, for the purpose of correctness, we need to construct a \defn{trivial isomorphism}
\begin{equation}\label{eq:triviso}
f : \BHil_j\otimes \BHil_k \to \BHil_j\otimes \BHil_{j-1},
\end{equation}
which is defined as the bijection between basis vectors with the same index.

Now we may consider two-qubit gates in their full generality. Let $U:\BHil_j\otimes \BHil_k\to \BHil_j\otimes \BHil_k$ be the gate under consideration. 
Perform a ``change of coordinates'' on $U$ to define \[V\coloneqq f U f^{-1}:\BHil_j\otimes \BHil_{j-1}\to \BHil_j\otimes \BHil_{j-1},\] and lift $V$ to $\tilde V : \Hil\to \Hil$ by way of \eqref{eq:tensortwo}. Then the lifted $U$ can be constructed as follows:
\begin{equation}
\tilde U = 
\begin{cases}
\pi_{j,k}^{-1}\tilde V\pi_{j,k} & \text{if }j > k\text{, and}\\
(\pi_{j,k}')^{-1}\tilde V\pi_{j,k}' & \text{if }j < k.
\end{cases}
\end{equation}
Since the $\pi$ operators are essentially compositions of involutive \SWAP{} gates, their inverses are just reversals.

With care, the essence of this method generalizes accordingly for gates acting on an arbitrary number of qubits.

We end this section with an example of a universal QAM.
\begin{example}\label{ex:universalqam}
Define all possible liftings of $U$ within $\Hil$ as
\begin{displaymath}
L(U) \coloneqq \{\text{$U$ lifted for all qubit permutations}\}.
\end{displaymath}
Define the gates \(\mathsf{S}\coloneqq
\left(
\begin{smallmatrix}
1 & 0\\
0 & i
\end{smallmatrix}
\right)
\) and \(\mathsf{T}\coloneqq
\left(
\begin{smallmatrix}
1 & 0\\
0 & e^{i\pi/4}
\end{smallmatrix}
\right)
\). A QAM with\footnote{When the context is clear, $G$ is sometimes abbreviated to just be the set of unlifted gates, but it should be understood that it's actually every lifted combination. See Section~\ref{sec:gates}.} \[G=L(\mathsf{H})\cup L(\mathsf{S})\cup L(\mathsf{T})\cup L(\CNOT) \text{ and } G'=\{\}\] can compute to arbitrary accuracy the action of any $N_q$-qubit gate, possibly with $P$ exponential in length. See \cite[\S4.5]{nielsen2010quantum} for details.
\end{example}

\subsection{Measurement Semantics}\label{sec:measurementsemantics}
Measurement is a surjective-only operation and is non-deterministic. In the space of a single qubit $Q_k$, there are two outcomes to a measurement. The outcomes are determined by lifting and applying---up to a scalar factor---either of the \defn{measurement operators}\footnote{In Dirac notation, $\bra{v}$ lies in the dual space of $\ket{v}$.}
\begin{equation*}
M_0^k \coloneqq \ket{0}_k\bra{0}_k \qquad\text{and}\qquad  M_1^k \coloneqq \ket{1}_k\bra{1}_k
\end{equation*}
to the quantum state. These can be interpreted as projections onto either of the basis elements of the Hilbert space. More generally, in any finite Hilbert space $\Hil$, we have the set of measurement operators \[M(\Hil) \coloneqq \big\{\ket{v}\bra{v} \mathbin{:} \ket{v}\in\operatorname{basis} \Hil\big\}.\] In the QAM, when a qubit is measured, a particular measurement operator $\mu$ is selected and applied according to the probability
\begin{equation}
P(\mu) \coloneqq P(\text{$\tilde\mu$ is applied during meas.}) = \bra{\Psi} \tilde\mu^{\dagger}\tilde\mu\ket{\Psi}.
\end{equation}
Upon selection of an operator, the quantum state transforms according to
\begin{equation}
\ket{\Psi} \leftarrow \frac{1}{P(\mu)}\tilde\mu\ket{\Psi}.
\end{equation}
This irreversible operation is called \defn{collapse of the wavefunction}.

In quantum mechanics, measurement is much more general than the description given above. In fact, any Hermitian operator can correspond to measurement. Such operators are called \defn{observables}, and the quantum state of a system can be seen as a superposition of vectors in the observable's eigenbasis. The eigenvalues of the observable are the outcomes of the measurement, and the corresponding eigenvectors are what the quantum state collapses to. For additional details on measurement and its quantum mechanical interpretation, we refer the interested reader to~\cite{nielsen2010quantum} and~\cite{aaronson2013quantum}.

\section{Quil: a Quantum Instruction Language}\label{sec:quil}
Quil is an instruction-based language for representing quantum computations with classical feedback and control. In its textual format---as presented below---it is line-based and assembly-like. It can be written directly for the purpose of quantum programming, used as an intermediate format within classical programs, or used as a compilation target for quantum programming languages. Most importantly, however, Quil defines what can be done with the QAM. Quil has support for:
\begin{itemize}
\item Applying arbitrary quantum gates,
\item Defining quantum gates as optionally parameterized complex matrices,
\item Defining quantum circuits as sequences of other gates and circuits, which can be bit- and qubit-parameterized,
\item Expanding quantum circuits,
\item Measuring qubits and recording the measurements into classical memory,
\item Synchronizing execution of classical and quantum algorithms,
\item Branching unconditionally or conditionally on the value of bits in classical memory, and
\item Including files as modular Quil libraries such as the standard gate library (see Appendix~\ref{sec:stdgates}).
\end{itemize}
By virtue of being instruction-based, Quil code effects transitions of the QAM as a state machine. In the next subsections, we will describe the various elements of Quil, using the syntax and conventions outlined in Section~\ref{sec:qam}.

\subsection{Classical Addresses and Qubits}
The central atomic objects on which various operations act are qubits, classical bits designated by an address, and classical memory segments.
\begin{description}
\item[Qubit] A qubit is referred to by its integer index. For example, $Q_5$ is referred to by \verb|5|.
\item[Classical memory address] A \defn{classical memory address} is referred to by an integer index in square brackets. For example, the address~7 pointing to the bit $C[7]$ is referred to as~\verb|[7]|.
\item[Classical memory segment] A \defn{classical memory segment} is a contiguous range of addresses from $a$ to $b$ inclusive with $a\le b$. These are written in square brackets as well, with a hyphen separating the range's endpoints. For example, the bits between 0 and 63 are addressed by~\verb|[0-63]| and represent the concatenation of bits \[C[63] C[62] \ldots C[1] C[0],\] written in the usual MSB-to-LSB order.
\end{description}

\subsection{Numerical Interpretation of Classical Memory Segments}
Classical memory segments can be interpreted as a numerical value for the purpose of controlling parametric gates. In particular, a 64-bit classical memory segment refers to an IEEE-754 double-precision floating point number \cite{ieee754}. A 128-bit classical memory segment $\texttt{[}x\texttt{-}(x+127)\texttt{]}$ refers to a double-precision complex number $a+ib$ where $a$ is the 64-bit interpretation of $\texttt{[}x\texttt{-}(x+63)\texttt{]}$, the first half of the segment, and $b$ is the 64-bit interpretation of $\texttt{[}(x+64)\texttt{-}(x+127)\texttt{]}$, the second half of the segment. The use of these numbers can be found in Section~\ref{sec:gates} and some practical consequences of their use can be found in Section~\ref{sec:VQE-dynamic}.

\subsection{Static and Parametric Gates}\label{sec:gates}
There are two gate-related concepts in the QAM: static and parametric gates. A \defn{static gate} is an operator in $\mathrm{U}(2^{N_q})$, and a \defn{parametric gate} is a function\footnote{Calling a parametric gate a ``gate'' is somewhat of a misnomer. The quantum gate is actually the \emph{image} of a point in $\mathbb{C}^n$.} $\mathbb{C}^n\to\mathrm{U}(2^{N_q})$, where the $n$ complex numbers are called \emph{parameters}. The implication is that \emph{operators in $G$ and $G'$ are always lifted to the Hilbert space of the QAM}. This is a formalism, however, and Quil abstracts away the necessity to be mindful of lifting gates.

In Quil, every gate is defined separately from its invocation. Each unlifted gate is identified by a symbolic name\footnote{To be precise, the symbolic name actually represents the equivalence class of operators under all trivial isomorphisms, as in \eqref{eq:triviso}.}, and is invoked with a fixed number of qubit arguments. The invocation of a static (resp.\ parametric) gate whose lifting is not a part of the QAM's $G$ (resp.\ $G'$) is undefined.

A static two-qubit gate named \verb|NAME| acting on $Q_2$ and $Q_5$, which is an operator lifted from $\BHil_2\otimes\BHil_5$, is written in Quil as the name of the gate followed by the qubit indexes it acts on, as in
\begin{verbatim}
NAME 2 5
\end{verbatim}
\begin{example}\label{ex:bell}
The Bell state on qubits $Q_0$ and $Q_1$ can be constructed with the following Quil code:
\begin{verbatim}
H 0
CNOT 0 1
\end{verbatim}
\end{example}

A parametric three-qubit gate named \verb|PNAME| with a single parameter $e^{-i\pi/7}$ acting on $Q_1$, $Q_0$, and $Q_4$, which is an operator lifted from $\BHil_1\otimes\BHil_0\otimes\BHil_4$, is written in Quil as
\begin{verbatim}
PNAME(0.9009688679-0.4338837391i) 1 0 4
\end{verbatim}
When a parametric gate is provided with a constant parameter, one could either consider the \emph{resulting} gate on the specified qubits to be a part of $G$, or the parametric gate itself on said qubits to be a part of $G'$.

Parametric gates can take a ``dynamic parameter'', as specified by a classical memory segment. Suppose a parameter is stored in \verb|[8-71]|. Then we can invoke the aforementioned gate with that parameter via
\begin{verbatim}
PNAME([8-71]) 1 0 4
\end{verbatim}
In some cases, using dynamic parameters can be expensive or infeasible, as discussed in Section~\ref{sec:VQE-dynamic}. Gates which use dynamic parameters are elements of $G'$.

\subsection{Gate Definitions}
Static gates are defined by their real or complex matrix entries in the basis described in Section~\ref{sec:qubitsemantics}. Matrix entries can be written literally with scientific \verb|E|-notation  (e.g., real \texttt{-1.2e2} or complex $\texttt{0.3-4.1e-4i}=0.3-0.00041i$), or built up from constant arithmetic expressions. These are:
\begin{itemize}
\item Simple arithmetic: addition \verb|+|, subtraction/negation \verb|-|, multiplication \verb|*|, division \verb|/|, exponentiation \verb|^|,
\item Constants: \verb|pi| (= $pi$), \verb|i| (= \verb|1.0i|), and
\item Functions: \verb|sin|, \verb|cos|, \verb|sqrt|, \verb|exp|, \verb|cis|\footnote{$\operatorname{cis}\theta \coloneqq \cos\theta + i\sin\theta=\exp{i\theta}$}.
\end{itemize}
The gate is declared using the \verb|DEFGATE| directive followed by comma-separated lists of matrix entries indented by exactly four spaces. Matrices that are not unitary (up to noise or precision) have undefined\footnote{Software processing Quil is encouraged to warn or error on such matrices.} execution semantics.

\begin{example}
The Hadamard gate can be defined by
\begin{verbatim}
DEFGATE HADAMARD:
    1/sqrt(2), 1/sqrt(2)
    1/sqrt(2), -1/sqrt(2)
\end{verbatim}
This gate is included in the collection of standard gates, but under the name \verb|H|.
\end{example}

Parametric gates are the same, except for the allowance of \emph{formal parameters}, which are names prepended with a `\texttt{\%}' symbol. Comma-separated formal parameters are listed in parentheses following the gate name, as is usual.
\begin{example}
The rotation gate $\RX$ can be defined by
\begin{verbatim}
DEFGATE RX(%theta):
    cos(%theta/2), -i*sin(%theta/2)
    -i*sin(%theta/2), cos(%theta/2)
\end{verbatim}
This gate is also included in the collection of standard gates.
\end{example}

Defining several gates or circuits with the same name is undefined.

\subsection{Circuits}
Sometimes it is convenient to name and parameterize a particular sequence of Quil instructions for use as a subroutine to other quantum programs. This can be done with the \verb|DEFCIRCUIT| directive. Similar to the \verb|DEFGATE| directive, the body of a circuit definition must be indented exactly four spaces. Critically, it specifies a list of \emph{formal arguments} which can be substituted with either classical addresses or qubits.

\begin{example}
In example~\ref{ex:bell}, we constructed a Bell state on $Q_0$ and $Q_1$. We can generalize this for arbitrary qubits $Q_m$ and $Q_n$ by defining a circuit.
\begin{verbatim}
DEFCIRCUIT BELL Qm Qn:
    H Qm
    CNOT Qm Qn
\end{verbatim}
With this, Example~\ref{ex:bell} is replicated by just a single line:
\begin{verbatim}
BELL 0 1
\end{verbatim}
\end{example}

Similar to parametric gates, \verb|DEFCIRCUIT| can optionally specify a list of parameters, specified as a comma-separated list in parentheses following the circuit name, as the following example shows.

\begin{example}
Using the $x$-$y$-$z$ convention, an extrinsic Euler rotation by $(\alpha, \beta, \gamma)$ of the state of qubit $q$ on the Bloch sphere is codified by the following circuit:
\begin{verbatim}
DEFCIRCUIT EULER(%alpha, %beta, %gamma) q:
    RX(%alpha) q
    RY(%beta)  q
    RZ(%gamma) q
\end{verbatim}
\end{example}

Within circuits, labels are renamed uniquely per expansion. As a consequence, it is possible to expand the same circuit multiple times, but it is not possible to jump into a circuit.

Circuits are intended to be used more as macros than as specifications for general quantum circuits. Indeed, \verb|DEFCIRCUIT| is very limited in its expressiveness, only performing argument and parameter substitution. It is included mainly to help with the debugging and human readability of Quil code.  More advanced constructs are intended to be written on top of Quil, as in Section~\ref{sec:tools}.

\subsection{Measurement}
Measurement provides the ``side effects'' of quantum programming, and is an essential part of most practical quantum algorithms (e.g., phase estimation and teleportation). Quil provides two forms of measurement: measurement-for-effect, and measurement-for-record.

Measurement-for-effect is a measurement performed on a single qubit used solely for changing the state of the quantum system. This is done with a \verb|MEASURE| instruction of a single argument. Performing a measurement on $Q_5$ is written as
\begin{verbatim}
MEASURE 5
\end{verbatim}
More useful, however, is measurement-for-record. Measurement-for-record is a measurement performed and recorded in classical memory. This form of the \verb|MEASURE| instruction takes two arguments, the qubit and the classical memory address. To measure $Q_7$ and deposit the result at address $8$ is written
\begin{verbatim}
MEASURE 7 [8]
\end{verbatim}
The semantics of the measurement operation are described in Section~\ref{sec:measurementsemantics}.

\begin{example}
Producing a random number between $0$ and $3$ inclusive can be accomplished with the following program:
\begin{verbatim}
H 0
H 1
MEASURE 0 [0]
MEASURE 1 [1]
\end{verbatim}
The memory segment \verb|[0-1]| is now representative of the number in binary.
\end{example}

\subsection{Program Control}\label{sec:programcontrol}
Program control is determined by the state of the program counter. The program counter $\kappa$ determines if the program has halted, and if not, it determines the location of the next instruction to be executed. Every instruction, except for the following, has the effect of incrementing $\kappa$. The exceptions are:
\begin{itemize}
\item Conditional and unconditional jumps.
\item The halt instruction \verb|HALT| which terminates execution and assigns $\kappa\leftarrow\vert P\vert$.
\item The last instruction in the program, which---after its execution---implicitly terminates execution as if by \verb|HALT|.
\end{itemize}
Locations within the instruction sequence are denoted by \defn{labels}, which are names that are prepended with an `\verb|@|' symbol, like \verb|@start|. The declaration of a new label within the instruction sequence is called a \defn{jump target}, and is written with the \verb|LABEL| directive.

Unconditional jumps are executed by the \verb|JUMP| instruction which sets $\kappa$ to the index of a given jump target.

Conditional jumps are executed by the \verb|JUMP-WHEN| (resp.\ \verb|JUMP-UNLESS|) instruction, which set $\kappa$ to the index of a given jump target if the bit at a classical memory address is $1$~(resp.\ $0$), and to $\kappa+1$ otherwise. This is a critical and differentiating element of Quil; it allows fast classical feedback.

\begin{example}
Consider the following C-like pseudocode of an \verb|if|-statement branching on the bit contained at address \verb|x|:
\begin{verbatim}
if (*x) {
    // instrA...
} else {
    // instrB...
}
\end{verbatim}
This can be translated into Quil in the following way:
\begin{verbatim}
JUMP-WHEN @THEN [x]
# instrB...
JUMP @END
LABEL @THEN
# instrA...
LABEL @END
\end{verbatim}
Lines starting with the \verb|#| character are comments and are ignored.
\end{example}

Labels that are declared within the body of a \verb|DEFCIRCUIT| are unique to that circuit. While it is possible to jump out of a \verb|DEFCIRCUIT| to a globally declared label, it is not possible to jump inside of one.
\begin{example}
Consider the following two \verb|DEFCIRCUIT| declarations and their instantiations. Note the comments on correct and incorrect usages of \verb|JUMP|.
\begin{verbatim}
DEFCIRCUIT FOO:
    LABEL @FOO_A
    JUMP @GLOBAL   # valid, global label
    JUMP @FOO_A    # valid, local to FOO
    JUMP @BAR_A    # invalid

DEFCIRCUIT BAR:
    LABEL @BAR_A
    JUMP @FOO_A    # invalid

LABEL @GLOBAL
FOO
BAR
JUMP @FOO_A        # invalid
JUMP @BAR_A        # invalid

\end{verbatim}
\end{example}

\subsection{Zeroing the Quantum State}
The quantum state of the QAM can be reset to the zero state with the \verb|RESET| instruction. This has the effect of setting \[\ket{\Psi}\leftarrow\bigotimes_{k=0}^{N_q-1}\ket{0}_{N_q-k-1}.\]
There are no provisions to clear the state of a single qubit, but we can do this by taking advantage of projective measurement.
\begin{example}
We can clear a qubit using a single bit of classical scratch space.
\begin{verbatim}
DEFCIRCUIT CLEAR q scratch_bit:
    MEASURE q scratch_bit
    JUMP-UNLESS @end scratch_bit
    X q
    LABEL @end
\end{verbatim}
\end{example}

\subsection{Classical/Quantum Synchronization}
Some classical/quantum programs can be constructed in a way such that at a certain point of a quantum program, execution must be suspended until some classical computations are performed and corresponding classical state is modified. This is accomplished using the \verb|WAIT| instruction, a synchronization primitive which signals to the classical computer that computation will not continue until some condition is satisfied. \verb|WAIT| takes no arguments.

The mechanism by which \verb|WAIT| works is deliberately unspecified. Some example mechanisms include monitors and interrupts, depending on the QAM implementation. An example use of \verb|WAIT| can be found in Section~\ref{sec:VQE-dynamic}.

\subsection{Classical Instructions}
Quil is intended to be a language to manipulate quantum state using quantum operations and classical control. Classical computation on the classical state should be done as much as possible with a classical computer, and using Quil's classical/quantum synchronization to mediate the hand-off of classical data between the classical and quantum processors. However, a few instructions for manipulating the classical state are provided for convenience, with emphasis on making control flow easier.

\subsubsection{Classical Unary Instructions}
The classical unary instructions are instructions that take a single classical address as an argument and modify the bit at that address accordingly. In each of the following, let $a$ be the address provided as the sole argument. The three instructions are:
\begin{description}
    \item[Constant False] \verb|FALSE|, which has the effect $C[a] \leftarrow 0$;
    \item[Constant True] \verb|TRUE|, which has the effect $C[a] \leftarrow 1$; and
    \item[Negation] \verb|NOT|, which has the effect $C[a] \leftarrow 1 - C[a]$.
\end{description}

\subsubsection{Classical Binary Instructions}
The classical binary instructions are instructions that take two classical addresses as arguments, and modify the bits at those addresses accordingly. In all of the following, let $a$ be the first address and $b$ be the second address provided as arguments. The four instructions are:
\begin{description}
    \item[Conjunction] \verb|AND|, which has the effect $C[b] \leftarrow C[a]C[b]$;
    \item[Disjunction] \verb|OR|, which has the effect \[C[b] \leftarrow 1 - (1 - C[a])(1 - C[b]);\]
    \item[Copy] \verb|MOVE|, which has the effect $C[b] \leftarrow C[a]$; and
    \item[Exchange] \verb|EXCHANGE|, which has the effect of swapping the bits at $a$ and $b$: $C[a]\leftrightarrow C[b]$.
\end{description}

\begin{example}
Exclusive disjunction $r\leftarrow a+b\mod 2$ can be implemented with the following circuit:
\begin{verbatim}
DEFCIRCUIT XOR a b r:
    # Uses (a | b) & (~a | ~b)
    MOVE b r
    OR a r              # r = a | b
    JUMP-UNLESS @end r  # short-circuit
    MOVE b r
    NOT a
    NOT r
    OR a r              # r = ~a | ~b
    NOT a               # undo change to a
    LABEL @end    
\end{verbatim}
Note that \verb|r| has to be distinct from \verb|a| and \verb|b|.
\end{example}

\subsection{The No-Operation Instruction}
The \defn{no-operation}, \defn{no-op}, or \verb|NOP| instruction does not affect the state of the QAM except in the way described in Section~\ref{sec:programcontrol}, i.e., by incrementing the program counter. This instruction may appear useless, especially on a machine with no notion of alignment or dynamic jumps. However, it has purpose when the QAM is used as the foundation for hardware emulation. For example, consider a QAM with some associated gate noise model. If one were to use an identity gate in place of a no-op, then the identity gate would be interpreted as noisy while the no-op would not. Moreover, the no-op has no qubit dependencies, which would otherwise affect program analysis. Rigetti Computing has used the no-op instruction as a way to force a break in instruction parallelization, described in Section~\ref{sec:parallel}.

\subsection{File Inclusion Semantics}
File inclusion is done via the \verb|INCLUDE| directive. For example, the library of standard gates---described in Appendix~\ref{sec:stdgates}---can be included with
\begin{verbatim}
INCLUDE "stdgates.quil"
\end{verbatim}
File inclusion is \emph{not} simple token substitution as it is in languages like the C preprocessor. Instead, the included file is parsed into a set of circuit definitions, gate definitions, and instruction code. Instruction code is substituted verbatim, but definitions will be recorded as if they were originally placed at the top of the file.

Generally, best practice is to include files containing \emph{only} contain gate or circuit definitions (in which case the file is called a \emph{library}), or \emph{only} executable code, and not both. However, this is not enforced.

\subsection{Pragma Support}
Programs that process Quil code may want to take advantage of extra information provided by the programmer. This is especially true when targeting QPUs where additional information about the machine's characteristics affect how the program will be processed. Quil supports a \verb|PRAGMA| directive to include extra information in a program which does not otherwise affect execution semantics. The syntax of \verb|PRAGMA| is as follows:
\begin{verbatim}
PRAGMA <identifier>+ <string>?
\end{verbatim}
where \verb|+| indicates one or more instances and \verb|?| indicates zero or one instance.
\begin{example}
The QAM does not have any notion of instruction parallelism, but programs are generally parallelized before execution on a QPU. (See Section~\ref{sec:parallel}.) Programs processing Quil may wish to enforce boundaries across which parallelization may not occur. An implementation may opt to support a parallelization barrier pragma. Despite the fact that the $\mathsf{X}$-gates below are commuting, an implementation may opt to treat the instructions sequentially.
\begin{verbatim}
X 0
PRAGMA parallelization_barrier
X 1
\end{verbatim}
Note that this does not change the semantics of Quil vis-\`a-vis the QAM.
\end{example}
\begin{example}
On modern superconducting qubit architectures, applications of different gates take different times. This can affect how instructions get scheduled for execution. Programs processing Quil may wish to allow the physical time it takes to perform a gate to be defined with a pragma, like so:
\begin{verbatim}
PRAGMA gate_time H "50 ns"
PRAGMA gate_time CNOT "150 ns"
H 0
CNOT 0 1
\end{verbatim}
\end{example}

\subsection{The Standard Gates}
Quil provides a collection of standard one-, two-, and three-qubit gates, which are fully defined in Appendix~\ref{sec:stdgates}. The standard gates can be used by including \verb|stdgates.quil|.

\section{Quil Examples}

\subsection{Quantum Fourier Transform}
In the context of the QAM's quantum state $\ket{\Psi}\in\Hil$, the \defn{quantum Fourier transform} (QFT) \cite[\S5.1]{nielsen2010quantum} is a unitary operator $\textsf{F}:\Hil\to\Hil$ defined by the matrix
\begin{equation}
\textsf{F}_{j,k} \coloneqq \frac{1}{\sqrt{2^{N_q}}}\omega^{jk},\qquad 0\le j,k < \dim\Hil = 2^{N_q}
\end{equation}
with $\omega\coloneqq \exp(2\pi i/2^{N_q})$ the complex primitive root of unity. It can be shown that this operator acts on the basis vectors (up to permutation) via the map
\begin{gather}
\bigotimes_{\mathclap{\substack{k=0\\k'\coloneqq N_q-k-1}}}^{N_q-1} \ket{b_{k'}}_{k'}
\mapsto
\frac{1}{\sqrt{2^{N_q}}}\bigotimes_{k=0}^{N_q-1}
\left[\ket{0}_{k} + \phi(k)\ket{1}_{k}\right],\\
\text{where } \phi(k) = \prod_{j=0}^k \exp\left(2\pi i b_{N_q-j-1}/2^{j+1}\right).\nonumber
\end{gather}
Here, $b$ is the bit string representation of the basis vectors, as in Section~\ref{sec:qubitsemantics}.

The first thing to notice is that the basis elements get reversed in this factorization. This is easily fixed via a product of \SWAP{} gates. In the context of classical fast Fourier transforms \cite{cooley1965algorithm}, this is called \defn{bit reversal}.

The second and more important thing to notice is that each factor of the $\bigotimes$ can be seen as an operation on the qubit $Q_{k}$. The factors of $\phi(k)$ indicate the operations are two-qubit \emph{controlled-phase} or \CPHASE{} gates with $Q_{k}$ as the target, and each previous qubit as the control. In the degenerate one-qubit case, this is a Hadamard gate. Further details on this algorithm can be found in~\cite{nielsen2010quantum}.

The core QFT logic can be implemented as a straightforward recursive function \verb|QFT'|. We write it as one which transforms qubits $Q_k$ for $l\le k<r$. The base case is the action on a single qubit---a Hadamard. In the general case, we do a sequence of \CPHASE{} gates targeted on the current qubit $Q_l$ and controlled by all qubits before it, topped off by a Hadamard. In the Python-like pseudocode below, we generate Quil code for this algorithm. We prepend lines of Quil code to be generated with two colons `\verb|::|'.
\begin{verbatim}
def QFT'(l, r):
    n = r - l        # num qubits
    if n == 1:       # base case
        :: H l
    else:            # general case
        QFT'(l + 1, r)
        for i in range(1, n):
            q = l + n - i
            alpha = pi / 2 ** (n - i)
            :: CPHASE(alpha) l q
        :: H l
\end{verbatim}
The bit reversal routine can be implemented straightforwardly as exactly $\lfloor N_q/2\rfloor$ \SWAP{} gates.
\begin{verbatim}
def revbin(Nq):
    for i in range(Nq / 2):
        :: SWAP i (Nq - i - 1)
\end{verbatim}
All of this is put together in a final \verb|QFT| routine.
\begin{verbatim}
def QFT(Nq):
    QFT'(0, Nq)
    revbin(Nq)
\end{verbatim}

\subsection{Variational Quantum Eigensolver}\label{sec:VQE}

The variational quantum eigensolver (VQE)~\cite{peruzzo2014variational, wecker2015progress, mcclean2015theory} is a classical/quantum algorithm for finding eigenvalues of a Hamiltonian $H$ variationally. The quantum subroutine has two fundamental steps.
\begin{enumerate}
\item Prepare the quantum state $\ket{\Psi(\vec\theta)}$, often called the \emph{ansatz}.
\item Measure the expectation value
$\langle\,\Psi(\vec\theta)\;|\,H\,|\;\Psi (\vec\theta)\,\rangle$.
\end{enumerate}
The classical portion of the computation is an optimization loop.
\begin{enumerate}
\item Use a classical non-linear optimizer to minimize the expectation value by varying ansatz parameters $\vec\theta$.
\item Iterate until convergence.
\end{enumerate}
We refer to given references for details on the algorithm.

Practically, the quantum subroutine of VQE amounts to preparing a state based off of a set of parameters $\vec\theta$ and performing a series of measurements in the appropriate basis. Using the QAM, these measurements will end up in classical memory. Doing this iteratively followed by a small amount of postprocessing, one may compute a real expectation value for the classical optimizer to use.

This algorithm can be implemented in at least two distinct styles which impose different requirements on a quantum computer.  These two styles can serve as prototypical examples for programming a QAM with Quil.

\subsubsection{Static Implementation}
One simple implementation of VQE is to generate a new Quil listing for every iteration of $\vec\theta$. Before calling out to the quantum subroutine, a new Quil program is generated for the next desired $\vec\theta$ and loaded onto the quantum computer for immediate execution. For a parameter $\theta=0.00724\ldots$, one such program might look like
\begin{verbatim}
# State prep
...
CNOT 5 6
CNOT 6 7
RZ(0.00724195969993) 7
CNOT 6 7
...
# Measurements
MEASURE 0 [0]
MEASURE 1 [1]
...
\end{verbatim}

This technique poses no issue from the perspective of coherence time, but adds a time penalty to each iteration of the classical optimizer.

A static implementation of VQE was written using Rigetti Computing's pyQuil library and the Rigetti QVM, both of which are mentioned in Sections~\ref{sec:tools} and \ref{sec:qvm} respectively.

\subsubsection{Dynamic Implementation}\label{sec:VQE-dynamic}
Perhaps the most encapsulated implementation of VQE would be to use dynamic parameters. Without loss of generality, let's assume a single parameter $\theta$. We can define a circuit which takes our single $\theta$ parameter and prepares $\ket{\Psi}$.
\begin{verbatim}
DEFCIRCUIT PREP_STATE(%theta):
    ...
    H 3
    H 5
    ...
    CNOT 3 5
    CNOT 5 6
    RZ(%theta) 6
    ...
\end{verbatim}
Next, we define a memory layout for classical/quantum communication:
\begin{itemize}
\item \verb|[0]|: Flag to indicate convergence completion.
\item \verb|[1-64]|: Dynamic parameter $\theta$.
\item \verb|[100]|, \verb|[101]|, \dots: Measurements corresponding to $Q_0$, $Q_1$, \dots.
\end{itemize}
Finally, we can define our VQE circuit.
\begin{verbatim}
DEFCIRCUIT VQE:
    LABEL @REDO
    RESET
    PREP_STATE([1-64]) # Dynamic Parameter
    MEASURE 0 [100]
    MEASURE 1 [101]
    ...
    WAIT
    JUMP-UNLESS @REDO [0]
\end{verbatim}

This program has the advantage that the quantum portion of the algorithm is completely encapsulated. It is not necessary to dynamically reload Quil code for each newly varied parameter.

The main disadvantage of this approach is its implementation difficulty in hardware. This is because of the diminished potential for program analysis to occur before execution. The actual gates that get applied will not be known until the runtime of the algorithm (hence the ``dynamic'' name). This may limit opportunities for optimization and poses particular issues for current quantum computing architectures which have limited natural gate sets and limited high-speed dynamic tune-up of new gates or their approximations.

\section{Forest: A Quantum Programming Toolkit}\label{sec:toolkit}
\subsection{Overview}
\begin{figure}
    \centering
    \includegraphics[scale=0.35]{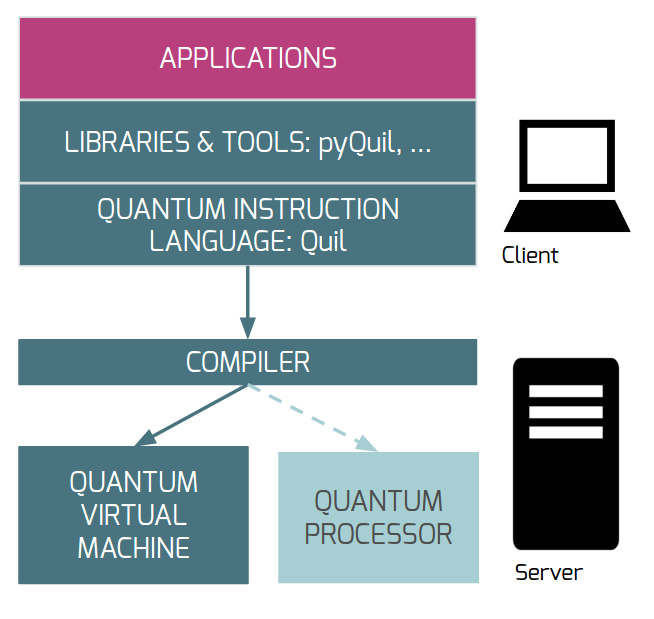}
    \caption{Outline of Forest, Rigetti Computing's quantum programming toolkit, described in Section~\ref{sec:toolkit}.}
    \label{fig:toolkit}
\end{figure}
Quantum computers, and specifically QAM-conforming QPUs, are not yet widely available. Despite this, software can make use of the the QAM and Quil to
\begin{enumerate*}[label={(\alph*)}]
\item study practical quantum algorithmic performance in a spirit similar to MMIX~\cite{KnutMMIX2005},
\item prepare a suite of software which will be able to be run on QPUs, and
\item provide a uniform interface to physical and virtualized QAMs.
\end{enumerate*}
Rigetti Computing has built a toolkit called \emph{Forest} for accomplishing these tasks.

The hierarchy of software roughly falls into four layers, as in figure~\ref{fig:toolkit}. Descending in abstraction, we have
\begin{description}
\item[Applications \& Tools] Software written to accomplish real-world tasks (e.g., study electronic structure problems), and software to assist quantum programming.
\item[Quil] The language described in this document and associated software for processing Quil. It acts as an intermediate representation of general quantum programs.
\item[Compiler] Software to convert arbitrary Quil to simpler Quil (or some other representation) for execution on a QPU or QVM.
\item[Execution Units] A QPU, a QVM, or a hardware emulator. Software written at this level will typically incorporate noise models intrinsic to a particular QPU.
\end{description}
We will briefly describe each of these components of the toolkit in the following sections.

\subsection{Applications and Tools}\label{sec:tools}
Quil is an assembly-like language that is intended to be both human readable and writable. However, more expressive power comes from being able to manipulate Quil programs programmatically. Rigetti Computing has developed a Python library called \emph{pyQuil}~\cite{pyQuil} which allows the construction of Quil programs by treating them as first-class objects. Using pyQuil along with scientific libraries such SciPy~\cite{SciPy}, Rigetti Computing has implemented non-trivial algorithms such as VQE using the abstractions of the QAM.

\subsection{Quil Manipulation}
Quil, as a language in its own right, is amenable to processing and computation independent of any particular (real or virtual) machine for execution. Rigetti Computing has written a reusable static analyzer and parser application for Quil, that allows Quil to easily be interchanged between programs. For example, Rigetti Computing's \verb|quil-json| program converts Quil into a structured JSON \cite{json} format:
\begin{verbatim}
$ cat bell.quil
H 0
CNOT 0 1
$ quil-json bell.quil
{
  "type": "parsed_program",
  "executable_program": [
    {
      "type": "unresolved_application",
      "operator": "H",
      "arguments": [["qubit", 0]],
      "parameters": null
    },
    {
       "type": "unresolved_application",
       "operator": "CNOT",
       "arguments": [["qubit", 0],
                     ["qubit", 1]],
       "parameters": null
        }
    ]
}
\end{verbatim}
Note the two instances of ``\texttt{unresolved\_application}''. These were generated because of a simple static analysis determining that these gates were not otherwise defined in the Quil file. (This could be ameliorated by including \verb|stdgates.quil|.)

\subsection{Compilation}\label{sec:compilation}
In the context of quantum computation, \defn{compilation} is the process of converting a sequence of gates into an approximately equivalent sequence of gates executable on a quantum computer with a particular qubit topology. This requires two separate kinds of processing: gate approximation and routing.

\begin{figure}[ht]
    \centering
    \includegraphics[scale=0.5]{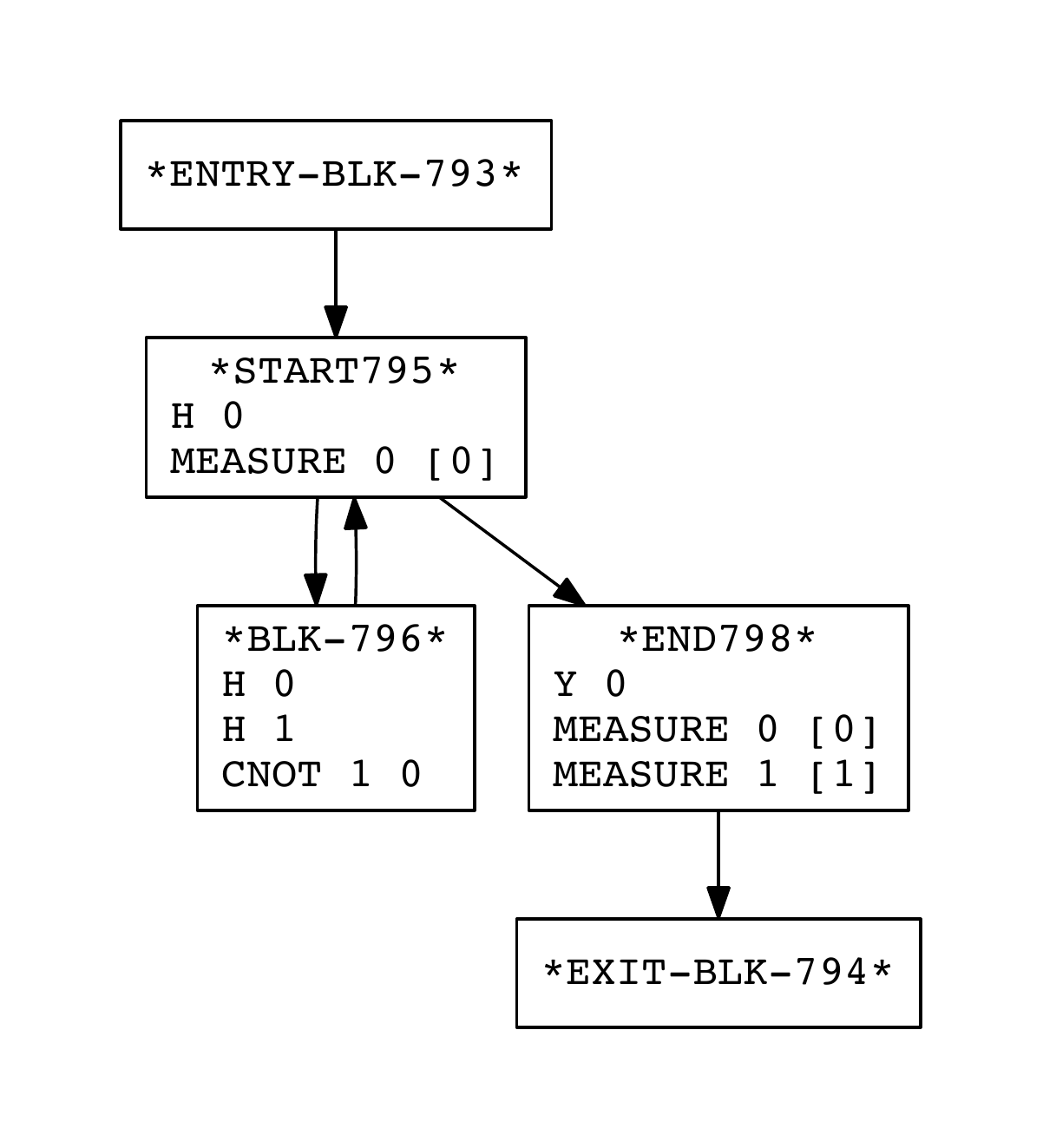}
    \caption{The control flow graph of a Quil program.}
    \label{fig:cfg}
\end{figure}

Since Quil is specified as a symbolic language, it is amenable to symbolic analysis. Quil programs can be decomposed into \emph{control flow graphs} (CFGs) \cite{Allen:1970:CFA:390013.808479} whose nodes are \emph{basic blocks} of straight-line Quil instructions, as is typical in compiler theory. Arrows between basic blocks indicate transfers of control. For example, consider the following Quil program:
\begin{verbatim}
LABEL @START
H 0
MEASURE 0 [0]
JUMP-WHEN @END [0]
H 0
H 1
CNOT 1 0
JUMP @START
LABEL @END
Y 0
MEASURE 0 [0]
MEASURE 1 [1]
\end{verbatim}
Roughly speaking, each segment of straight-line control makes up a basic block. Figure~\ref{fig:cfg} depicts the control flow graph of this program, with jump instructions elided.

Many of these basic blocks will be composed of gates and measurements, which themselves can be symbolically and algebraically manipulated. Gates can go through \defn{approximation} to reduce a general set of gates to a sequence of gates native to the particular architecture, and then \defn{routing} so that these simpler gates are arranged to act on neighboring qubits in a physical architecture. Another example of a transformation on basic blocks is parallelization, talked about in the next section.

Both approximation and routing can be formalized as transformations between QAMs. In this first example, we show how we can formally describe a transformation between a QAM to another one with a smaller but computationally equivalent gate set.
\begin{example}[Compiling]\label{ex:compile}
Let $\mathfrak{M}=(\ket{\Psi}, C, G_{\mathfrak{M}}, G', P, \kappa)$ be a one-qubit QAM with \[G_{\mathfrak{M}}=\{\HADAMARD, \RX(\theta), \RZ(\theta)\}\] for some fixed $\theta\in\mathbb{R}$, and let $\mathfrak{M}'$ be a QAM with \[G_{\mathfrak{M}'}=\{\HADAMARD, \RZ(\theta)\}.\] Because $\RX = \HADAMARD\RZ\HADAMARD$, we can define a compilation function $\mathfrak{M}\mapsto\mathfrak{M}'$ specifically transforming\footnote{In the parlance of functional programming, $f$ is applied to $P$ via a \emph{concatmap} operation.} $\iota\in P$ according to
\begin{equation*}
    f(\iota) =
    \begin{cases}
    (\HADAMARD, \RZ(\theta), \HADAMARD) &\text{if $\iota=\RX(\theta)$,}\\
    (\iota)                             &\text{otherwise.}
    \end{cases}
\end{equation*}
\end{example}

In this next example, we show how qubit connectivity can be encoded in a QAM, and how one can route instructions to give the illusion of a fully connected topology.

\begin{example}[Routing]
Let $\mathfrak{M}=(\ket{\Psi}, C, G_{\mathfrak{M}}, G', P, \kappa)$ be a three-qubit QAM with
\[G_{\mathfrak{M}}=L(\HADAMARD)\cup L(\CNOT)\cup L(\SWAP),\]
where $L$ was defined in Example~\ref{ex:universalqam}. Consider a three-qubit QPU with the qubits arranged in a line \[\textrm{$Q_0$---$Q_1$---$Q_2$}\] so that two-qubit gates can only be applied on adjacent qubits. Then this QPU can be modeled by another three-qubit QAM $\mathfrak{M}'$ with the lifted gates
\[G_{\mathfrak{M}'}=
\left\{
\begin{array}{c}
\HADAMARD_0, \HADAMARD_1,  \HADAMARD_2,\\
\CNOT_{01}, \CNOT_{10}, \CNOT_{12}, \CNOT_{21},\\
\SWAP_{01}, \SWAP_{12}
\end{array}
\right\}.\]
Because of the qubit topology, there are no gates that act on $\BHil_2\otimes\BHil_0$. However, we can reason about transforming between $\mathfrak{M}$ and $\mathfrak{M}'$ in either direction. We can transform from $\mathfrak{M}$ to $\mathfrak{M}'$ by way of a transformation $f$ similar to that in the last example, namely
\begin{equation*}
    f(\iota) =
    \begin{cases}
    (\SWAP_{01}, \SWAP_{12}, \SWAP_{01}) &\text{if $\iota=\SWAP_{02}$,}\\
    (\SWAP_{01}, \CNOT_{12}, \SWAP_{01}) &\text{if $\iota=\CNOT_{02}$,}\\
    (\SWAP_{01}, \CNOT_{21}, \SWAP_{01}) &\text{if $\iota=\CNOT_{20}$,}\\
    (\iota)                              &\text{otherwise.}
    \end{cases}
\end{equation*}
Similarly, we can transform from $\mathfrak{M}'$ to $\mathfrak{M}$ by adding three additional gates to $G_{\mathfrak{M}'}$, namely those implied by $f$.
\end{example}

Many other classes of useful transformations on QAMs exist, such as $G$-preserving algebraic simplifications of $P$, an analog of peephole optimization in compiler theory \cite{McKeeman:1965:PO:364995.365000}.

\subsection{Instruction Parallelism}\label{sec:parallel}
\defn{Instruction parallelism}, the ability to apply commuting operations to a quantum state in parallel, is one of the many benefits of quantum computation. Quil code as written is linear and serial, but can be interpreted as an instruction-parallel program. In particular, many subsequences of Quil instructions may be executed in parallel. Such sequences include:
\begin{itemize}
\item Commuting gate applications and measurements, and
\item Measurements with non-overlapping memory addresses.
\end{itemize}
In general, parallelization cannot occur over jumps, resets, waits, and measurements and dynamic gate applications with overlapping address ranges. We suggest that \verb|NOP| is used as a way to force a parallelization break.

\begin{figure}[ht]
    \centering
    \includegraphics[scale=0.5]{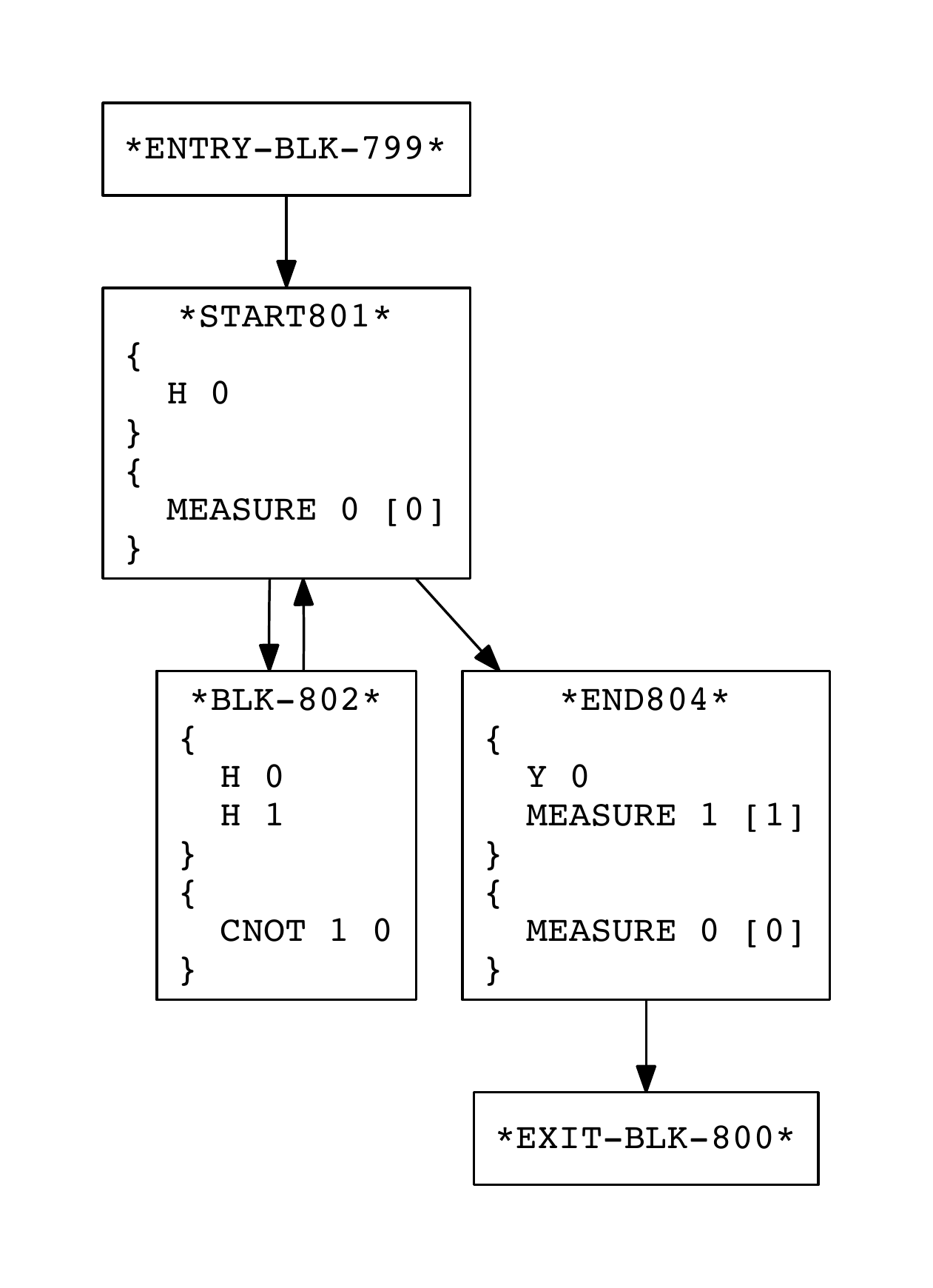}
    \caption{The parallelized version of Figure~\ref{fig:cfg}. Instruction sequences which can be executed in parallel are surrounded in curly braces `\texttt{\{\}}'.}
    \label{fig:cfg-parallel}
\end{figure}

If a control flow graph is constructed as in Section~\ref{sec:compilation}, then parallelization can be done over each basic block. A parallelized basic block is called a \defn{parallelization schedule}. See Figure~\ref{fig:cfg-parallel} for an example of quantum parallelization within a CFG.

\subsection{Rigetti Quantum Virtual Machine}\label{sec:qvm}
Rigetti Computing has implemented a QVM in ANSI Common Lisp~\cite{ANSI:1996:ANS} called the \emph{Rigetti QVM}. It is a high-performance\footnote{The Rigetti QVM has optimized vectorized and parallelized numerics, and has no theoretical limit for the number of qubits it can simulate. It has been demonstrated to simulate 36 qubits.}, programmable simulator for the QAM and emulator for noisy quantum computers. The Rigetti QVM exposes two interfaces for executing quantum programs: execution of Quil files directly with POSIX-style shared memory (``local execution''), and execution of Quil from HTTP server requests (``remote execution'').

Local execution is useful for high-speed testing of small-to-medium sized instances of classical/quantum algorithms. It also provides convenient ways of debugging quantum programs, such as allowing direct inspection of the quantum state at any point in the execution process, as well as limited quantum hardware emulation with tunable noise models.

Remote execution is used for distributed, cloud access to QVM instances. HTTP libraries exist in nearly all modern programming languages, and allow programs in these languages to make connections. Rigetti Computing has built in to pyQuil the ability to send the first-class Quil program objects to a local or secured remote Rigetti QVM instance using the Forest API.

\section{Conclusion}
We have introduced a practical abstract machine for reasoning about and executing quantum programs. Furthermore, we have described a notation for such programs on this machine, which is amenable to analysis, compilation, and execution. Finally, we have described a pragmatic toolkit for quantum program construction and execution built atop these ideas.

\section{Acknowledgements}
The authors would like to thank their colleagues at Rigetti Computing for their support, especially Nick Rubin. We are also grateful to Erik Davis, Jarrod McClean, Bill Millar, and Eric Peterson for their helpful discussions and valuable suggestions provided throughout the development of this work.

\appendix

\subsection{The Standard Gate Set}\label{sec:stdgates}
The following static and parametric gates are defined in \verb|stdgates.quil|. Many of these gates are standard gates used in theoretical quantum computation \cite{nielsen2010quantum}, and some of them find their origin in the theory of superconducting qubits \cite{chow2010quantum}.
\begin{description}
\item[Pauli Gates]
\begin{align*}
\texttt{I} &= \left(\begin{smallmatrix}
1 & 0\\
0 & 1
\end{smallmatrix}\right)
&
\texttt{X} &= \left(\begin{smallmatrix}
0 & 1\\
1 & 0
\end{smallmatrix}\right)
&
\texttt{Y} &= \left(\begin{smallmatrix}
0 & -i\\
i & 0
\end{smallmatrix}\right)
&
\texttt{Z} &= \left(\begin{smallmatrix}
1 & 0\\
0 & -1
\end{smallmatrix}\right)
\end{align*}
\item[Hadamard Gate]
\begin{displaymath}
\texttt{H} = \tfrac{1}{\sqrt{2}}\left(\begin{smallmatrix}
1 & 1\\
1 & -1
\end{smallmatrix}\right)
\end{displaymath}
\item[Phase Gates]
\begin{align*}
\texttt{PHASE}(\theta) &= \left(\begin{smallmatrix}
1 & 0\\
0 & e^{i\theta}
\end{smallmatrix}\right)
&
\texttt{S} &= \texttt{PHASE}(\pi/2)
&
\texttt{T} &= \texttt{PHASE}(\pi/4)
\end{align*}
\item[Controlled-Phase Gates]
\begin{align*}
\texttt{CPHASE00}(\theta) &= \diag(e^{i\theta},1,1,1) \\
\texttt{CPHASE01}(\theta) &= \diag(1,e^{i\theta},1,1) \\
\texttt{CPHASE10}(\theta) &= \diag(1,1,e^{i\theta},1) \\
\texttt{CPHASE}(\theta) &= \diag(1,1,1,e^{i\theta})
\end{align*}
\item[Cartesian Rotation Gates]
\begin{align*}
\texttt{RX}(\theta) &= \left(\begin{smallmatrix}
\cos\frac{\theta}{2} & -i\sin\frac{\theta}{2}\\
-i\sin\frac{\theta}{2} & \cos\frac{\theta}{2}
\end{smallmatrix}\right)\\
\texttt{RY}(\theta) &= \left(\begin{smallmatrix}
\cos\frac{\theta}{2} & -\sin\frac{\theta}{2}\\
\sin\frac{\theta}{2} & \cos\frac{\theta}{2}
\end{smallmatrix}\right)\\
\texttt{RZ}(\theta) &= \left(\begin{smallmatrix}
e^{-i\theta/2} & 0\\
0 & e^{i\theta/2}
\end{smallmatrix}\right)
\end{align*}
\item[Controlled-$\mathsf{X}$ Gates]
\begin{align*}
\texttt{CNOT} &=
\left(
\begin{smallmatrix}
1 & 0 & 0 & 0\\
0 & 1 & 0 & 0\\
0 & 0 & 0 & 1\\
0 & 0 & 1 & 0
\end{smallmatrix}
\right) &
\texttt{CCNOT} &=
\left(
\begin{smallmatrix}
    1 & 0 & 0 & 0 & 0 & 0 & 0 & 0\\
    0 & 1 & 0 & 0 & 0 & 0 & 0 & 0\\
    0 & 0 & 1 & 0 & 0 & 0 & 0 & 0\\
    0 & 0 & 0 & 1 & 0 & 0 & 0 & 0\\
    0 & 0 & 0 & 0 & 1 & 0 & 0 & 0\\
    0 & 0 & 0 & 0 & 0 & 1 & 0 & 0\\
    0 & 0 & 0 & 0 & 0 & 0 & 0 & 1\\
    0 & 0 & 0 & 0 & 0 & 0 & 1 & 0
\end{smallmatrix}
\right)
\end{align*}
\item[Swap Gates]
\begin{align*}
\texttt{PSWAP}(\theta) &=
\left(
\begin{smallmatrix}
1 & 0 & 0 & 0\\
0 & 0 & e^{i\theta} & 0\\
0 & e^{i\theta} & 0 & 0\\
0 & 0 & 0 & 1
\end{smallmatrix}
\right)\\
\texttt{SWAP} &= \texttt{PSWAP}(0)\\
\texttt{ISWAP} &= \texttt{PSWAP}(\pi/2)\\
\texttt{CSWAP} &=
\left(
\begin{smallmatrix}
    1 & 0 & 0 & 0 & 0 & 0 & 0 & 0\\
    0 & 1 & 0 & 0 & 0 & 0 & 0 & 0\\
    0 & 0 & 1 & 0 & 0 & 0 & 0 & 0\\
    0 & 0 & 0 & 1 & 0 & 0 & 0 & 0\\
    0 & 0 & 0 & 0 & 1 & 0 & 0 & 0\\
    0 & 0 & 0 & 0 & 0 & 0 & 1 & 0\\
    0 & 0 & 0 & 0 & 0 & 1 & 0 & 0\\
    0 & 0 & 0 & 0 & 0 & 0 & 0 & 1
\end{smallmatrix}
\right)
\end{align*}
\end{description}

\subsection{Prior Work}\label{sec:prior}
There exists much literature on abstract models, syntaxes, and semantics of quantum computing. Most of them are in the form of quantum programming languages (QPLs) and simulators, which achieve various levels of expressiveness. Languages roughly fall into three categories:
\begin{itemize}
    \item embedded domain-specific languages,
    \item high-level quantum programming languages, and
    \item low-level quantum intermediate representations.
\end{itemize}
In addition, work has been done on designing larger tool chains for quantum programming \cite{abhari2012scaffold,javadiabhari2015scaffcc,svore2006layered}.

In the following subsections, we provide a non-exhaustive account of some previous work within the above categories, and describe how they relate to the design of our quantum ISA.

\subsubsection{Embedded Domain-Specific Languages}
An \defn{embedded domain-specific language} (EDSL) is a language to express concepts and problems from within another programming language. Within the context of quantum programming, two prominent examples are Quipper~\cite{Quipper}, which is embedded in Haskell~\cite{Haskell}, and \Liquid{}~\cite{Liquid}, which is embedded in F\#~\cite{FSharp}. Representation of quantum programs (and, in particular, the subclass of quantum programs called \emph{quantum circuits}) is expressed with data structures in the host language. Feedback of classical information is achieved through an interface with the host language.

Quantum programs written in an EDSL are not directly executable on a quantum computer, due to the requirement of being present in the host language's runtime. However, since quantum programs are represented as first-class objects, these objects are amenable to processing and compilation to a quantum computer's natively executable instruction set. Quil is one such target for compilation.

\subsubsection{High-Level QPLs}

High-level QPLs are the quantum analog of languages like C~\cite{kandr} in the imperative case or ML~\cite{stdml} in the functional case. They provide a plethora of classical and quantum data types, as well as control flow constructs. In the case of functional quantum languages, the lambda calculus for quantum computation with classical control by Selinger and Valiron can act as a theoretical basis for their semantics of such a language\footnote{This is similar to how the classical untyped lambda calculus formed the basis for LISP in 1958~\cite{mccarthy1960recursive}.}~\cite{selinger2006lambda}.

One prominent example of a high-level QPL is Bernhard \"Omer's QCL~\cite{OmerB:strqp}. It is a C-like language with classical control and quantum data. Among the many that exist, two important---and indeed somewhat dual---data types are \verb|int| and \verb|qureg|. The following example shows a Hadamard initialization on eight qubits, and measuring the first four of those qubits into an integer variable.
\begin{verbatim}
// Allocate classical and quantum data
qureg q[8];
int m;
// Hadamard initialize
H(q);
// Measure four qubits into m
measure q[0..3], m;
// Print m, which will be anywhere from 0 to 15.
print m;
\end{verbatim}

\"Omer defines the semantics of this language in detail in his PhD thesis, and presents an implementation of a QCL interpreter written in C++~\cite{qclweb}. However, a compilation or execution strategy on quantum hardware is not presented. Similar to EDSLs, QPLs such as QCL can be compiled into a lower-level quantum instruction set such as Quil.

\subsubsection{Low-Level Quantum Intermediate Representations}
In the context of compiler theory, an \defn{intermediate representation} (IR) is some representation of a language which is amenable to further processing. IR can be higher level (as with abstract syntax trees), or lower level (as with linear bytecodes). Low-level IRs often act as compilation targets for classical programming languages, and are used for analysis and program optimization. For example, LLVM IR \cite{LLVM:CGO04} is used as an intermediate compilation target for the Clang C compiler~\cite{clang:URL}.

The most well-known example of a low-level quantum IR is QASM \cite{circqasm}. This was originally a language to describe the quantum circuits for \LaTeX{} output in \cite{nielsen2010quantum}, and hence, \emph{not} an IR in that form. However, the syntax was adapted for executable use in \cite{qasmtools}. QASM, however, does not have any notion of classical control within the language, acting solely as a quantum circuit description language.

Quil is considered a low-level quantum IR with classical control.

\bibliographystyle{IEEEtran}
\bibliography{references}
\end{document}